**Thermoluminescence Properties of Ce - doped and Ce/Tb-doped Sol-gel Silica Rods. Application to Passive (OSL) and Real-time (RL) Dosimetry**

Prof. M. Benabdesselam*, Dr. J. Bahout, Dr. F. Mady, Dr. W. Blanc, Dr. H. El Hamzaoui, Dr. A. Cassez, K. Delplace-Baudelle, R. Habert, Prof. G. Bouwmans, Prof. M. Bouazaoui, Prof. B. Capoen

Université Côte d'Azur, Institut de Physique de Nice (INPHYNI), 06108 Nice Cedex 2, France
E-mail : Mourad.Benabdesselam@univ-cotedazur.fr

Univ-Lille, CNRS, UMR 8523 – PhLAM – Physique des Lasers Atomes et Molécules, F-59000 Lille, France

**Keywords**

silica, rare earth, thermoluminescence, radioluminescence, optically stimulated luminescence, dosimetry

**Abstract**

Two rods made from sol-gel silica have been doped with Ce ions or co-doped with Ce and Tb ions respectively. First, a thermoluminescence (TL) characterization of the trapping and luminescence parameters is carried out in order to understand the physical mechanisms involved following the irradiation of these rods and to evaluate their dosimetric properties. The optically stimulated luminescence (OSL) and the radioluminescence (RL) responses as a function of respectively absorbed dose and dose rate are assessed. The OSL response of both rods is linear as a function of the absorbed dose. The RL sensitivity of both rods proves not only to be extremely high but also shows a linear behavior over more than 6 decades, allowing real-time detection of dose rates as low as few $\mu Gy\ s^{-1}$, a threshold that to our knowledge, has never been reached. The results from RL and OSL show that these silica-based doped rods are potentially suitable for medical dosimetry and environmental monitoring around nuclear sites.

# 1. Introduction

The characterization of a radiative medium is done by measuring the dose or the dose rate present in this medium. The detectors used for these measurements are based on two technologie: one electronic, the basic physical principle of which is radioelectric conversion (ionization chamber, diamond, diodes, MOSFETs, etc.) and the other is optical, essentially exploiting the properties of scintillation or luminescence of certain materials under irradiation. In this work, we are interested in the latter, having regard to some of its advantages in terms of electromagnetic immunity [1], luminescence properties for the assessment of the dose and/or dose-rate of irradiation. These properties are studied using well-known techniques such as thermoluminescence (TL), radioluminescence (RL), or even a more recent technique based on optical stimulation (OSL). The latter is a dosimetry technique whose physical principle is identical to that of TL but with an "all-optical" approach since the information on the dose contained in the dosimeter is collected by stimulation by means of a laser diode rather than by heating like in TL [2].

For luminescence dosimetry applications, the most widely used technique for a long time is undoubtedly that of TL using a variety of wide bandgap insulating materials, most often crystalline [3] but also amorphous [4,5] sensitive to the radiative environment, whether nuclear, spacial or medical. For example, in the medical field, the European Society for Radiotherapy and Oncology (ESTRO) uses TL dosimetry for the external control of radiotherapy machines [6].

During the 1980s, the advent of high-power stimulation electroluminescent diodes at different wavelengths combined with appropriate filtering made the OSL technique as another dosimetry tool [7] with a major advantage over TL since it does not require any energy input to heat the material up to a few hundred °C. However, the physics underlying the processes of TL and of OSL is exactly the same. Indeed, these two phenomena involve the same charge carrier trapping centers and the same recombinant ones. The only difference lies in the process of release of the trapped charges following irradiation. A wide range of materials already used as efficient TL dosimeters, have also been used in OSL dosimetry [1,8-11] but the most prevalent being probably $Al_2O_3$: C [12].

Whereas the two previous stimulation techniques are tools of choice for estimating the dose once the irradiation has ended, the RL, which consists of the spontaneous luminescence observed following the prompt recombination of the radiation-induced electron-hole pairs in material during irradiation, provides a real-time assessment of the dose rate. The materials commonly used in RL dosimetry are plastic scintillators [13] or doped inorganic crystals [14].

All of these materials exhibit luminescent properties and are all sensitive to the dose or dose-rate of radiation. However, their major drawback is that unlike glassy materials, they cannot be drawn in the form of optical fibers.

This study shows that the dose and dose-rate responses of doped vitreous silica rods produced at low cost and easily drawn into optical fibers make them potentially interesting in dosimetric monitoring. Indeed, they can be deployed over long distances in difficult-to-access areas like nuclear storage facilities [15] and also, given their small size, to increase spatial resolution in zones with a high dose gradient such as in IMRT (Intensity Modulated Radiation Therapy) [16]. Authors [17-19] have already reported some works on dosimetry by RL, OSL or both techniques on doped sol-gel materials. In this present work, we propose to compare to compare the dosimetric performances by OSL and by RL of two sol-gel silica rods, one simply Ce-doped and the other Ce/Tb co-doped. The two doping ions ($Ce^{3+}$ and $Tb^{3+}$) were chosen on the basis of their emission quantum efficiency in the visible domain once inserted in silica.

## 2. Experimental Section

### 2.1. Studied rods

Pure silica xerogels were synthesized by the sol-gel route from tetraethylorthosilicate (TEOS) precursor within the FiberTech platform in Lille. The obtained xerogels were stabilized at 1000°C and then soaked in alcoholic doping solution containing cerium and/or terbium salt. The samples were dried at 50°C to remove solvents and to retain doping species with the nanopores. The doped samples were then densified in helium atmosphere, as described elsewhere [20-21]. The obtained glasses were drawn to a millimeter-sized rods at a temperature of about 2000°C. Hence, the obtained samples are colorless doped silica glass rods with a diameter of about 1.6 mm. Their optical quality seems similar to that of the undoped specimen. Two rods were synthesized: the first one is Ce-doped and the second one is co-doped with both Ce and Tb. Contents of these doping elements are estimated between 200 and 300 at-ppm.

### 2.2. Experimental setup (irradiation, TL, OSL RL)

X-ray irradiations have been performed in air at room temperature (RT). The tube accelerates electrons towards a Cu target operating at a voltage of 30 kV (XRG3D – Inel). For the different doses and dose rates in silica ($SiO_2$), the irradiation beam is calibrated by means of a UnidosE electrometer connected to an ionization chamber (PTW 23342) placed in the isocenter of the beam where the studied rods should be positioned.

The TL signal of the rod placed on a heating planchet was recorded at a linear heating rate of 1 °C $s^{-1}$ between RT and 450 °C by means of a photomultiplier tube (PMT). The emission integration is spanned over a 250–650 nm spectral range, the delay between the irradiation and the beginning of the TL reading is usually around 1 minute. The distribution of the trapping levels was obtained using the Initial Rise (IR) method [22].

The RL/OSL double detection system is shown in **Figure 1**. It is made up of a photon counting PMT (H7421-40) and a block with fiber outputs enclosing a dichroic mirror. The photon counting PMT with a radiant peak sensitivity around 500 nm and a high signal-to-noise ratio, is more suitable for detection of the low light levels associated with the OSL and RL signals from materials exposed to low doses or dose rates. A 5 m long multimode silica transport fiber (SMMF) is spliced on one side to the silica rod to be irradiated and is connected to the dichroic mirror block on the other.

The OSL was recoded at RT using the setup shown in **Figure 1**. The detection of the luminescence was achieved through optical filters with transmission centered on the luminescence band of the rods previously irradiated. Optical stimulation was performed using a 660 nm laser diode (LD) of 60 mW output power. As in TL, the delay between the end of the irradiation and the laser stimulation was around 1 minute. After acquirement, rods were heated up to 500°C to remove any residual signal. The dose estimate is then obtained by calculating the integrated area of the OSL decay curve and subtracting the background noise.

RL measurements were conducted in the same setup except that this time the LD was off. Note that for RL and OSL measurements, the irradiations were carried out through a lead screen provided with a slit having the size of the rods so as to protect the transport fiber from the irradiation beam (**Figure 1**).

The spectral distribution of TL peaks was also measured from RT to 450 °C but by means of an optical multichannel analyzer (Fergie, Princeton Instruments), consisting of a multimode optical fiber connected to an integrated spectrograph equipped with a CCD array covering the range 200-1100 nm with high efficiency in the UV-Visible range.

## 3. Résultats and Discussion

### 3.1. TL features: trapping and recombination centers

The TL signal arises from thermally stimulated electron-hole recombination as trapped charge carriers are excited from their traps by the absorption of heating. energy. This measurement helps identifying how deep, with respect to the conduction or valence band, are the trapping centers that contribute to the OSL and RL signals by allowing an estimate of the traps' thermal stability.

**Figure 2** represents the normalized TL curves of the two rods, under the same conditions of irradiation and readout. The Ce-doped sample shows two peaks with approximately the same intensity, the first relatively thin at 67 °C and another much wider around 195 °C. The co-doped one presents also two main peaks. However, the first appears as a shoulder peaking at 67 °C and the second, larger and more intense at 166 °C. The peak at 67 °C, common to both rods, is always present in the TL glow curve of silica-based materials regardless of the dopant. It is probably due to the intrinsic OD (oxygen deficient) centers widely present in the silica matrix [23]. The peak at around 200 °C is induced by Ce, as it is often the case with RE doping (Tb, Eu, Yb, …) while the one located at 166 °C is introduced by the Ce/Tb co-doping. This shift towards low temperatures was also observed recently in the TL behavior of Ce/Cu co-doped samples [24].

In terms of phosphorescence, just at the start of the thermograms, the Ce-doped rod shows a more intense signal than the co-doped one. So, it seems that co-doping results in two effects: first, the decrease in the intensities of both phosphorescence and TL first peak would mean a decrease in the population of shallow trap levels; second, the shift of the main peak towards low temperatures is synonymous with a rearrangement of atoms in the matrix.

**Figure 3** shows the distribution of the trap levels within the bandgap of doped silica, estimated using the IR technique from RT up to $T_{\text{stop}} = 225$ °C. Unsurprisingly, as with amorphous materials, we obtain a continuous distribution of levels ranging between 0.9 and 1.5 eV. Up to about 140 °C, the repartition of the trap levels seems to be the same on both rods, meaning that the associated defects are probably intrinsic to silica. Above 140 °C, the distributions of the trap depths begin to be distinguished between the two rods.

The phosphorescence and TL peaks observed in **Figure 2** correspond respectively to the recombination of released carriers from the traps located at around 0.9-1 and 1.3-1.4 eV.

The representative spectrally-resolved TL of the rods all over the temperature range is presented in **Figure 4**. TL of Ce-doped rod consists in a large emission band centered at 470 nm (2.64 eV). The latter was already assigned to the recombination of trapped electrons involving the $5d^1 \rightarrow 4f^1$ transition of $Ce^{3+}$ ions during the stimulation process [25,26]. The TL emission spectrum of Ce/Tb co-doped rod is clearly a superposition of that of $Ce^{3+}$ ions and $Tb^3$ emission bands. $Tb^3$ ions emission is characterized by blue and green lines, due to transitions from respectively the $^5D_3$ and $^5D_4$ levels to $^7F_J$ [27,28]. To confirm this superposition, we added the TL emission spectrum of a Tb-doped rod. It is thus easy to notice that the spectrum of the Ce/Tb co-doped rod is, at least qualitatively, the superposition of the TL emissions of the rods single doped: Ce on the one hand and Tb on the other. This could also mean that there is no preferential spatial correlation between the trapping defects in the host matrix and the doping ions.

A possible mechanism to describe the TL process of the rods would be that initially, during irradiation, the electrons are captured by the various traps whose levels are between 0.9 and 1.4 eV (**Figure 3**) below the conduction band while the holes are captured on the doping ions. Then, during thermal stimulation, the released electrons recombine with the holes to give $(Ce^{3+})^*$ and $(Tb^{3+})^*$ ions in the excited state. The so-excited ions then emit at their characteristic wavelengths during a radiative relaxation to the ground state (**Figure 4**). The spectral resolution of the RL shows, at least qualitatively, the same emissions as in TL for all rods suggesting that the same recombination centers, i.e., the dopant ions are the luminescence centers in both processes.

It is worth to note that these blue and blue-green emissions of the rods are very suitable for the dosimetry by luminescence whether it is TL, RL, or OSL because they all fall within the spectral window of most commercial PMTs used in dosimetry.

### 3.2. OSL and RL dosimetry properties

*3.2.1. OSL dosimetry*

**Figures 5** represent the OSL response as a function of the absorbed dose obtained on the two rods, following X-ray irradiations at a same dose-rate of 0,11 Gy(SiO$_2$) s$^{-1}$.

At least up to absorbed doses of 100 Gy(SiO$_2$), the Ce-doped sample exhibits very good linearity in its OSL response. The signal of the Ce/Tb co-doped rod also shows a linear behavior but only up to an absorbed dose of 40 Gy(SiO$_2$), beyond which this response becomes sub-linear and tends towards saturation at higher doses. Up to absorbed doses of 40 Gy(SiO$_2$) for which the responses are both linear, we find that the OSL of the Ce-doped sample is at least twice as sensitive as that of the cod-doped one. response. This difference between the responses of the two rods could be due to a higher concentration of traps involved in the OSL process in the case of the Ce-doped rod, as it also appears to be indicated by comparing the areas under the TL curves (**Figure 2**).

Note that for both samples, the OSL signals are very reproducible and for a given dose, no dose-rate dependency was observed, at least between 0.1 and 10 Gy(SiO$_2$) s$^{-1}$, which can be useful in dosimetry.

*3.2.2. RL dosimetry*

In **Figure 6** is shown the normalized RL behavior as a function of time for the two rods at different dose rates. It is useful to specify that in the case of our 30 keV X-ray irradiation conditions, the measured RL signals are not altered by the stem effect, which is known to affect the length of the transport fiber exposed to the irradiation field [29]. Moreover, this transport fiber is protected from the radiation beam by a thick lead shield, as shown in **Figure 1**. As shown in the case of the co-doped rod (**Figure 6.b**), the measurements are very reproducible on both rods.

Looking at the form of the signals, either during irradiation or in its absence, it is easy to see that in the case of the Ce/Tb co-doped sample, the RL signal seems to suddenly reach stable levels as soon as the irradiation is turned ON, contrary to what is observed in the case of the Ce-doped rod, where much longer exposure times are needed to reach stable signal levels. An almost similar behavior is also observed when the irradiations are turned off, since the return to the background level of the RL signal is done with a certain latency, more pronounced in the case of the simply doped rod than in the co-doped one.

In brief, two phenomena seem to occur in the RL response: the first one at the time of irradiation initiation and the second one at its termination.

Concerning the first phenomenon appearing when the irradiation is ON, which is characterized by a more or less slow rise of the RL signal over time, it is probably due to the competition between the process of electrons trapping on the shallow traps and the direct recombination on the radio-luminescent centers. The "plateau" level without any influence of the shallow traps on the RL signal, is thus reached when the electrons' detrapping rate reaches that of their trapping. This phenomenon, more evident in the response of the Ce-doped rod than in the Ce/Tb co-doped one, agrees with **Figure 2** showing higher TL intensity of the peak at 67 °C on the Ce-doped rod than on the co-doped one. Hence, it suggests that this this TL peak at 67 °C is characteristic of these shallow traps involved in the RL process.

These shallow traps are also at the origin of the second phenomenon of slow signal decay at the end of the irradiation. This decay is in fact phosphorescence due to the electrons released from the shallow traps by the simple effect of the thermal stimulation at RT. Here again, the TL curves of **Figure 2** show a higher level of phosphorescence in the case of the Ce-doped rod, which could explain the difference in the decay to the background level between the two samples. The observations made on the shape of the RL signals and their evolution for the two rods are rigorously the same over a wide range of dose rates.

**Figure 7** shows RL responses as a function of the dose rate of the two rods. Both responses are reported to be linear over a wide dose-rate range extending on six decades [13 µGy(SiO$_2$) s$^{-1}$ – 10 Gy(SiO$_2$) s$^{-1}$] and probably more, if the X-ray generator was not limited.

Such a linearity range of the RL response is in very good agreement with the results obtained by Vedda et al. [18] on Ce-doped silica samples in similar X-ray irradiation conditions but only over three decades [0,1 mGy s$^{-1}$ – 50 mGy s$^{-1}$]. To our knowledge, a linearity of the response over such a wide range of dose rates has never been explored, especially in real time and in a fiber mode dosimetry configuration.

As noticed before, in the case of the Ce/Tb co-doped rod, stable RL levels are immediately achieved during irradiation, which is not the case with the Ce-doped sample. Thus, we have verified for the Ce-doped rod that the readings of the RL signal intensity at the middle of the irradiation or at its end lead to the same linear evolutions as those shown in **Figure 7a**. This means that, even in its still evolving form, the RL signal of Ce-doped rod might enable a quite reliable assessment of the dose-rate during the irradiation.

We just note that whatever the dose-rate considered; the Ce/Tb co-doping seems to provide enhanced RL light output with a sensitivity about one order of magnitude greater than that of the Ce-doped one. This suggests that dose-rates of the order of µGy(SiO$_2$) s$^{-1}$ remain easily measurable especially with the co-doped Ce/Tb rod.

The high sensitivity means that the dosimeters can be made small, which in turn gives them the property of high spatial resolution, meaning that they have the potential for measurement of dose in regions of severe dose gradients. The all-optical nature of the process means that they can be used with optical fibers to measure doses in difficult-to-access locations.

## 4. Conclusion

The TL study of two rare-earth doped sol-gel silica rods revealed many features at the origin of the RL and OSL mechanisms observed respectively during and after irradiation. Indeed, the different TL peaks are related to the trapping centers that constitute the source of the charge carriers giving rise to the OSL signal while the emissions of these peaks correspond to the radio-luminescent centers at the origin of the RL signal. Moreover, the TL signal at RT and the low temperature peaks, originating from the shallow traps account for the evolution kinetics of the RL signal at the start and at the end of the irradiation.

After TL, the OSL and RL responses of the rods were then investigated. It was shown that these fiber rods could be suitable for fiber dosimetry either by OSL or RL even if each of them shows a more or less a betters performance depending on the dosimetry technique considered. The Ce-doped sample seems more adapted to OSL dosimetry, given the greater sensitivity of its dose response in addition to the wider linearity range of this response. On the other hand, due to the rapidity at which the RL signal reaches stability and due to its higher sensitivity (about one order of magnitude), the Ce/Tb co-doped rod is the most suitable for real-time dosimetry, even though detection of dose rate levels as low as a few µGy s$^{-1}$ is achievable with both rods. Such detected low dose-rate levels show that the use of these fiber rods in RL could prove to be attractive for environmental monitoring of gamma radiation around nuclear storage sites. This attractiveness is even higher when considering that such rods could be easily drawn into fibers and thus deployed in several locations.

In the medical field, the RL responses of both specimens increase linearly over a very wide range of dose-rates, including those encountered in radiotherapy (10 -100 mGy s$^{-1}$). In addition, the Ce-doped rod exhibits a linear response of its OSL signal in absorbed dose range commonly used in radiotherapy (0.5-80 Gy). However, at the high energies of the medical field (around 10 MeV), as well as under gamma radiation around nuclear sites, the presence of the stem effect should not be obscured in the precise assessment of the RL response of these rods.

**Acknowledgements**

This operation was supported by Andra within the French government "Investissements d'Avenir" program: SURFIN Project (Andra) under grant #RTSCNADAA17-0044. It was also partially supported by the ANR: LABEX CEMPI (ANR-11-LABX-0007) and the Equipex Flux (ANR-11-EQPX-0017), by the Ministry of Higher Education and Research, the Hauts-de-France Regional Council and the European Regional Development Fund (ERDF) through the Contrat de Projets Etat-Region (CPER Photonics for Society P4S). This work has been also supported by IRCICA institute and FiberTech Lille platform of University of Lille.

**Conflict of interest**

The authors declare no conflict of interest

[1] A. L. Huston, B. L. Justus, P. L. Falkenstein, *Nucl. Instrum. Methods, Phys. Res. B* **2001**, *184*, 55.

[2] B. L. Justus, S. Rychnovsky, M. A. Miller, K. J. Pawlovich, A. L. Huston, *Radiat. Prot. Dosim*. **1997**, *74*, 151.

[3] S. W. S. McKeever, M. Moscovitch, D. Townsend, Thermoluminescence Dosimetry Materials: Properties and Uses, NTP, **1995**.

[4] A. L. Yusoff, R. P. Hugtenburg, D. A. Bradley, *Radiat. Phys. Chem.* **2005**, *74*, 459.

[5] M. Benabdesselam, F. Mady, S. Girard, Y. Mebrouk, J.B. Duchez, M. Gaillardin, P. Paillet, *IEEE Trans. Nucl. Sci.* **2013**, *60*, 4251.

[6] I. H. Ferreira, A. Dutreix, A. Bridier, J. Chavaudera, H. Svensson, *Radiotherapy and Oncology*, **2000**, *55*, 273.

[7] L. Botter-Jensen, S. W. S. McKeever, A. G. Wintle, Optically Stimulated Luminescence Dosimetry, Elsevier Science B. V., **2003**

[8] M. Sommer, J. Henniger, *Radiat. Prot. Dosim*. **2006**, *119*, 394.

[9] D. Lapraz, H. Prevost, K. Idri, G. Angellier, L. Dusseau, *Phys. Status Solidi a* **2006**, *203*, 3793.

[10] G. Espinosa, J. S. Bogard, *J. Radioan. Nucl. Chem.* **2008**, *277*, 125.

[11] B. L. Justus, S. Rychnovsky, M. A. Miller, K. J. Pawlovich and A. L. Huston, *Radiat. Prot. Dosim*. **1997**, *74*, 151.

[12] B. G. Markey, L. Botter-Jensen, N. R. J. Poolton, H. E. Christiansen, F. Willumsen, *Radiat. Prot. Dosim*. **1996**, *66*, 413.

[13] Syed Naeem Ahmed, Physics and Engineering of Radiation Detection, Elsevier, **2015**

[14] M. Nikl and A. Yoshikawa, *Adv. Opt. Mater*. **2015**, *3*, 463

[15] B. G. Risch, B. Overton, J. Rosko, A. Bergonzo, G. Kuyt and G. Mélin, *Proceedings of the 61st IWCS Conference*. **2012**, 543.

[16] P.C. Williams, *Br. J. Radiol*. **2003**, *76*, 766.

[17] N. Al Helou, H. El Hamzaoui, B. Capoen, G. Bouwmans, A. Cassez, Y. Ouerdane, A. Boukenter, S. Girard, G. Chadeyron, R. Mahiou and M. Bouazaoui, *IEEE Trans. Nucl. Sci.*, **2018**, *65*, 1591.

[18] A. Vedda, N. Chiodini, D. Di Martino, M. Fasoli, S. Keffer, A. Lauria, M. Martni, F. Moretti, G. Spinolo, M. Nikl and N. Slovieva, *Appl. Phys. Lett.* **2004**, *85*, 6356.

[19] B. Capoen, H. El Hamzaoui, M. Bouazaoui, Y. Ouerdane, A. Boukenter, S. Girard and O. Duhamel, *Opt. Mat.* **2016**, *51,* 104.

[20] H. El Hamzaoui, L. Courtheoux, V. Nguyen, E. Berrier, A. Favre, L. Bigot, M. Bouazaoui, B. Capoen, *Mater. Chem. Phys.* **2010**, *121*, 83.

[21] H. El Hamzaoui, G. Bouwmans, B. Capoen, Y. Ouerdane, G. Chadeyron, R. Mahiou, S. Girard, A. Boukenter, M. Bouazaoui, *Mater. Res. Express* **2014**, *1*, 026203-1.

[22] N. S. Rawat, M. S. Kulkarni, D. R. Mishra, B. C. Bhatt, C. M. Sunta, S. K. Gupta, D. N. Sharma, *Nucl. Instr. Method Phys. Res. B* **2009**, *267*, 3475.

[23] F. Mady, A. Gutilla, M. Benabdesselam, W. Blanc, *Opt. Mat. Express.* **2019**, *9*, 2466.

[24] M. Benabdesselam, F. Mady, A. Gutilla, W. Blanc, H. El Hamzaoui, M. Bouazaoui, N. Al Helou, J. Bahout, G. Bouwmans, B. Capoen, *IEEE Trans. Nucl. Sci.* **2020**, *67*, 1663.

[25] N. Chiodini, M. Fasoli, M. Martini, E. Rosetta, G. Spinolo and A. Vedda, *Appl. Phys. Lett.* **2002**, *81*, 4374.

[26] H. El Hamzaoui, B. Capoen, N. Al Helou, G. Bouwmans, Y. Ouerdane, A. Boukenter, S. Girard, C. Marcandella, O. Duhamel, G. Chadeyron, R. Mahiou, M. Bouazaoui, *Mater. Res. Express* **2016**, *3*, 046201-1.

[27] A. Vedda, N. Chiodini, D. Di Martino, M. Fasoli, L. Griguta, F. Moretti, E. Rosetta, *J. Non-Cryst. Solids* **2005**, *351*, 3699.

[28] M. Fasoli, F. Moretti, A. Lauria, N. Chiodini, A. Vedda, M. Nikl, *Rad. Meas.* **2007**, *42*, 784.

[29] B. L. Justus, P. Falkenstein, A. Huston, M. C. Plazas, H. Ning and R. Miller, *Appl. Opt.* **2004**, *65*, 1591.

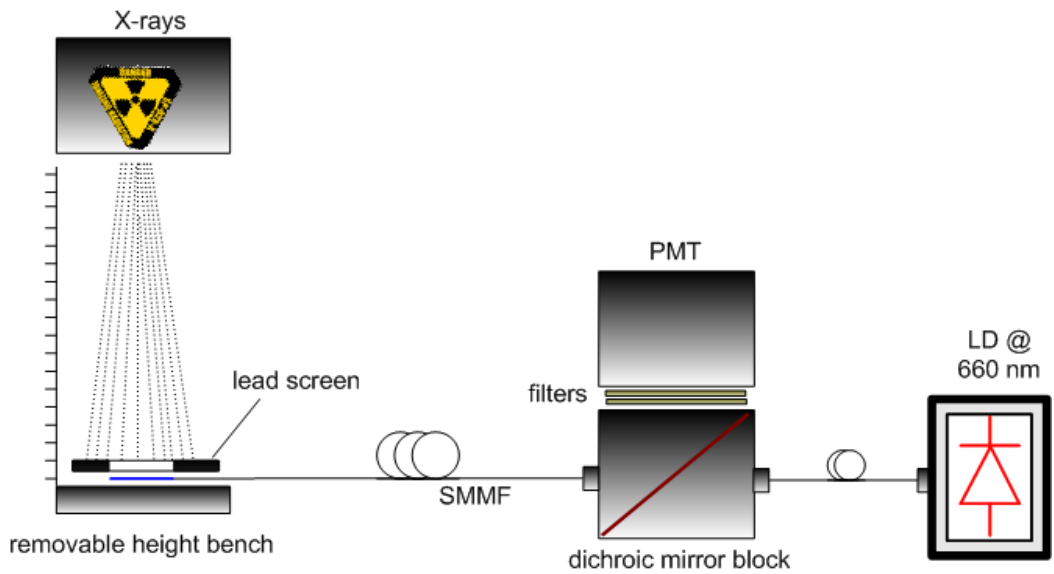

**Figure 1.** The RL/OSL double detection fiber system and the irradiation setup

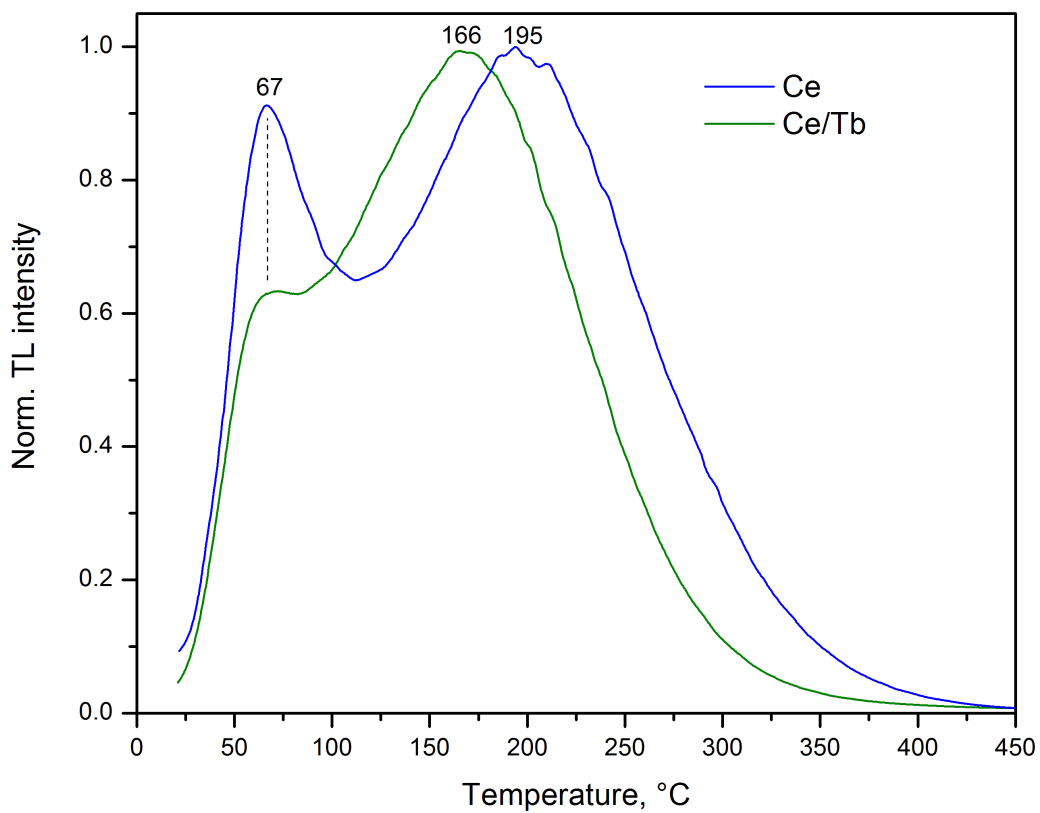

**Figure 2**. TL of Ce- and Ce/Tb doped sol-gel silica rods after X-ray irradiation at RT

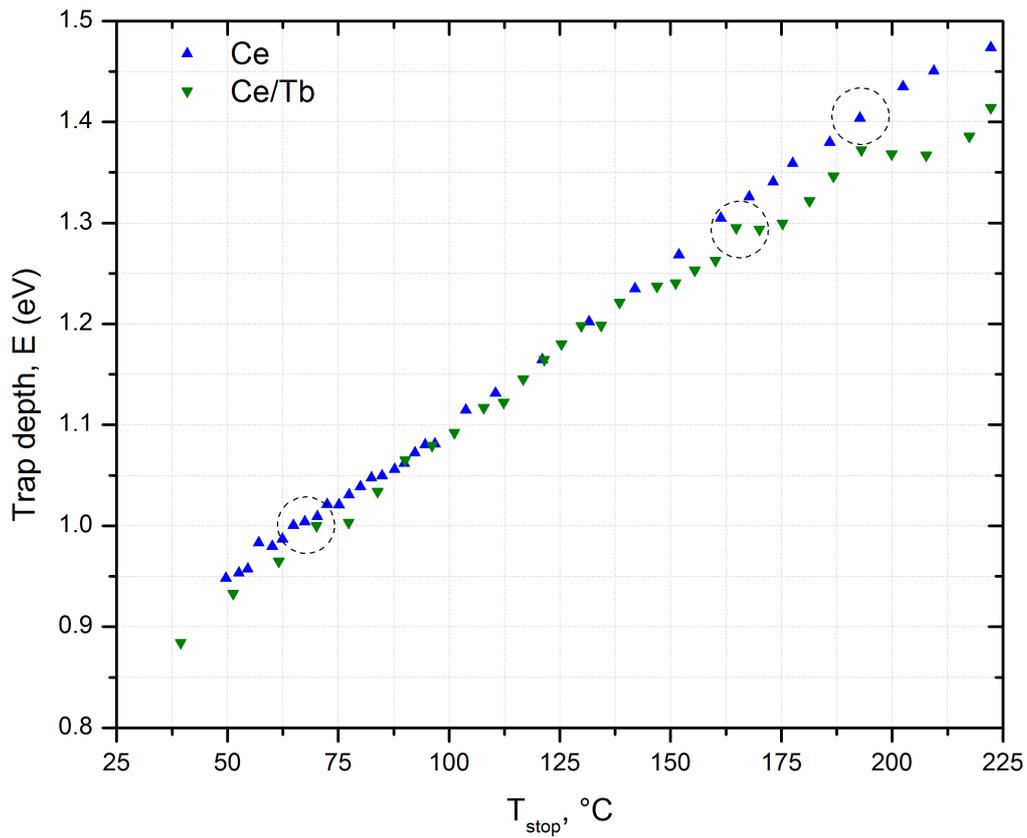

**Figure 3**. Trapping levels distribution of Ce- and Ce/Tb-doped sol-gel silica rods

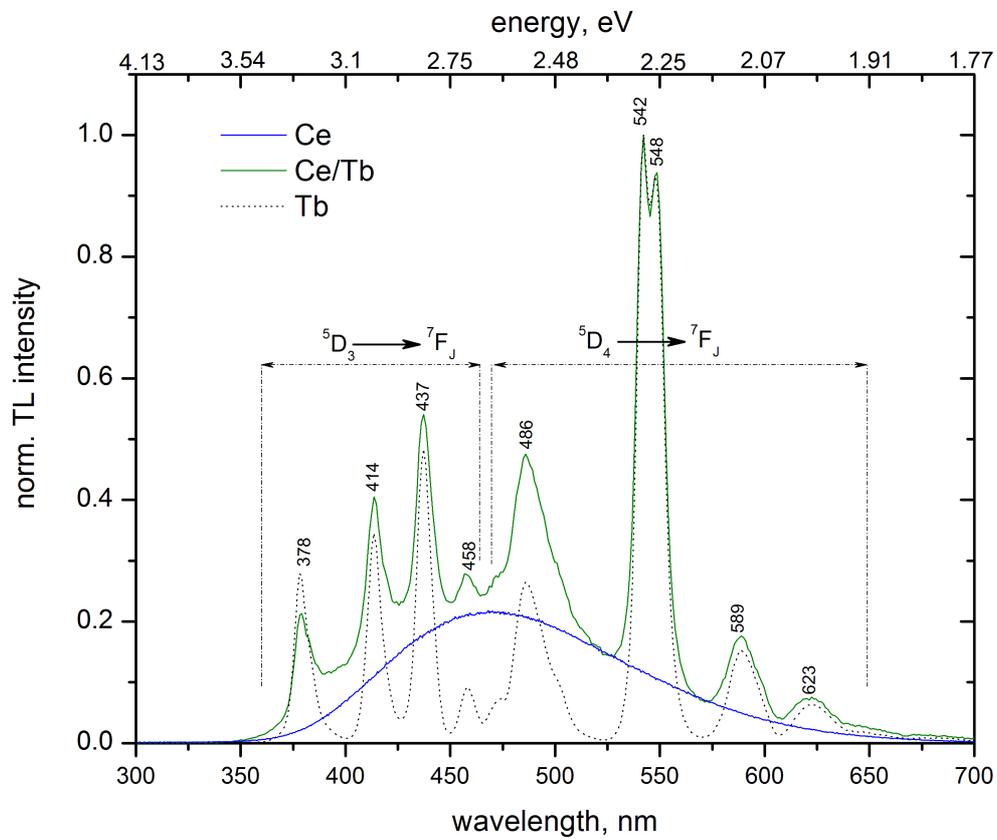

**Figure 4**. TL spectral distribution of Tb-, Ce- and Ce/Tb-doped sol-gel silica rods

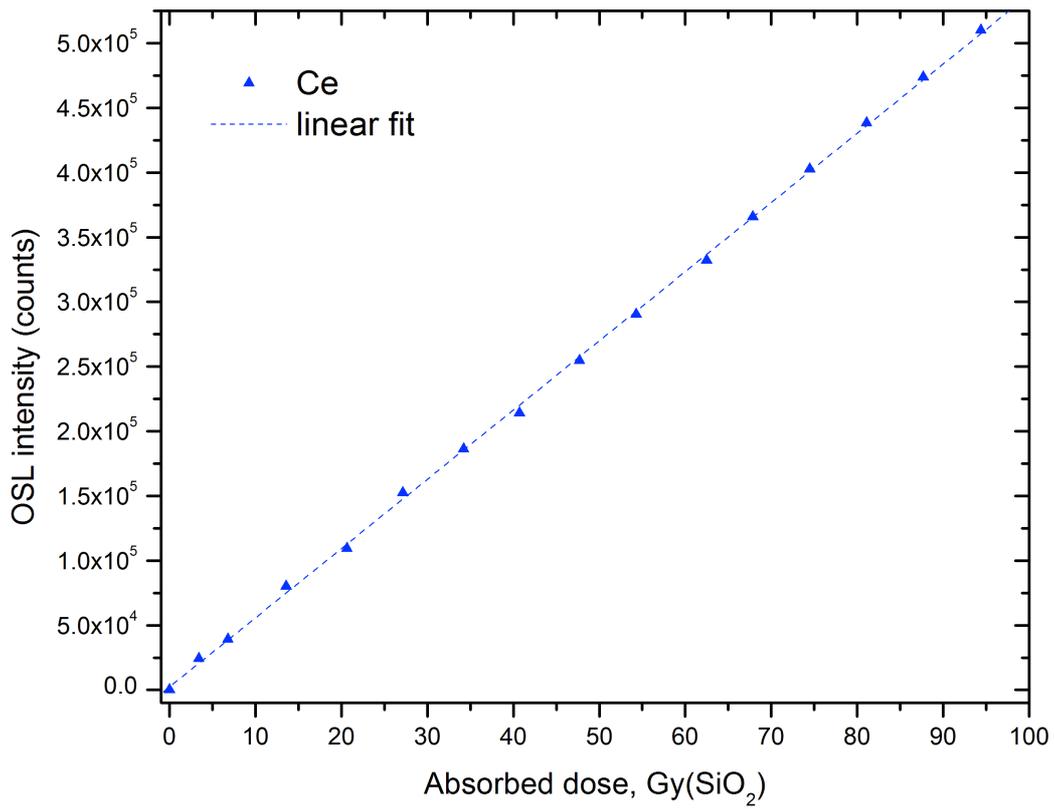

**Figure 5a.** Integrated OSL signal as a function of the absorbed dose of the Ce-doped rod.

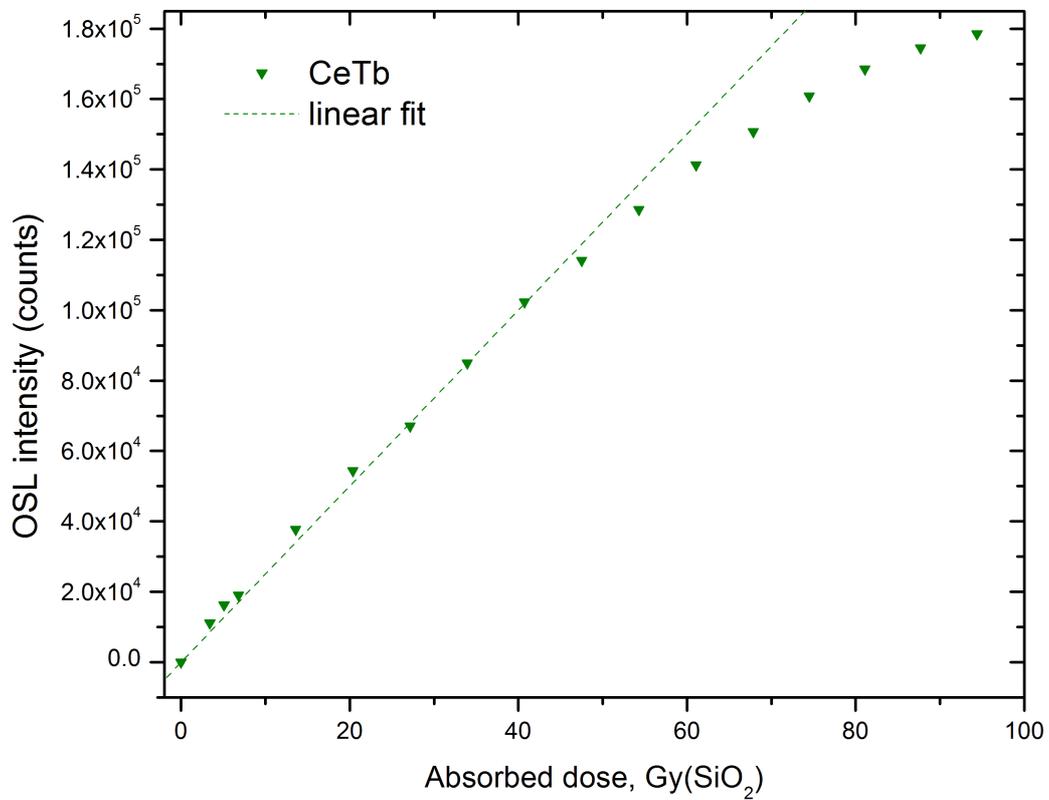

**Figure 5b.** Integrated OSL signal as a function of the absorbed dose of the Ce/Tb-doped rod.

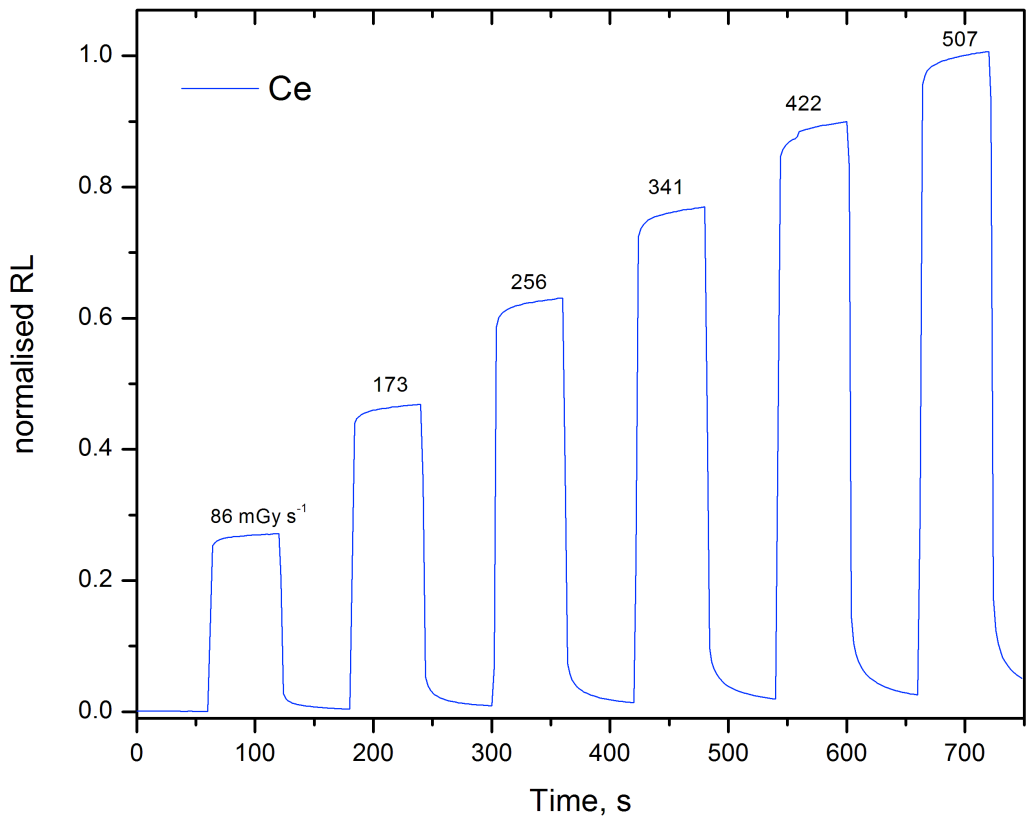

**Figure 6a**. Norm. RL behavior vs. the increasing dose-rate for Ce-doped rod.

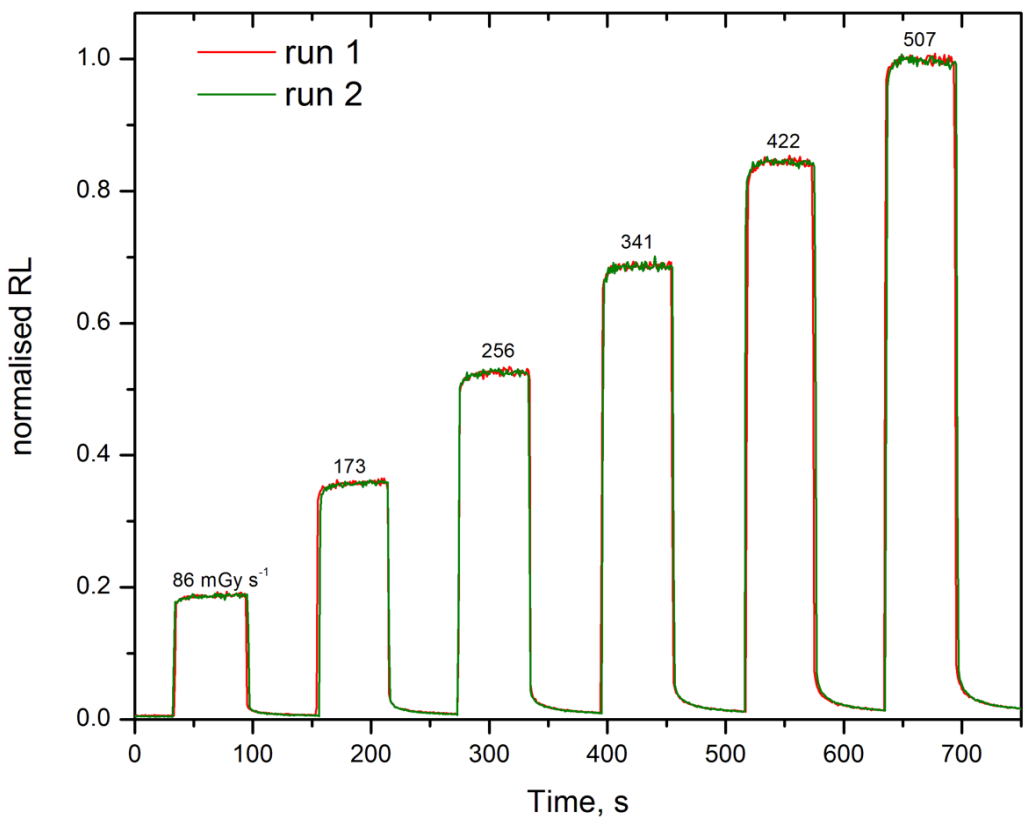

**Figure 6b**. Norm. RL behavior vs. the increasing dose-rate for Ce/Tb co-doped rod.

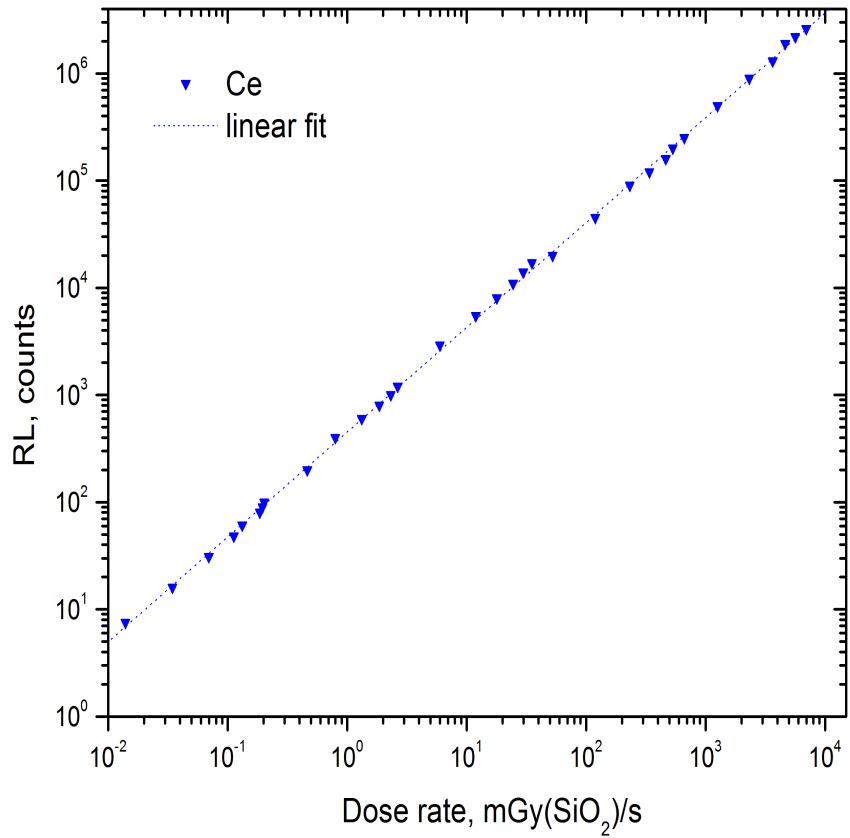

**Figure 7a**. Dose-rate dependence for Ce-doped rod

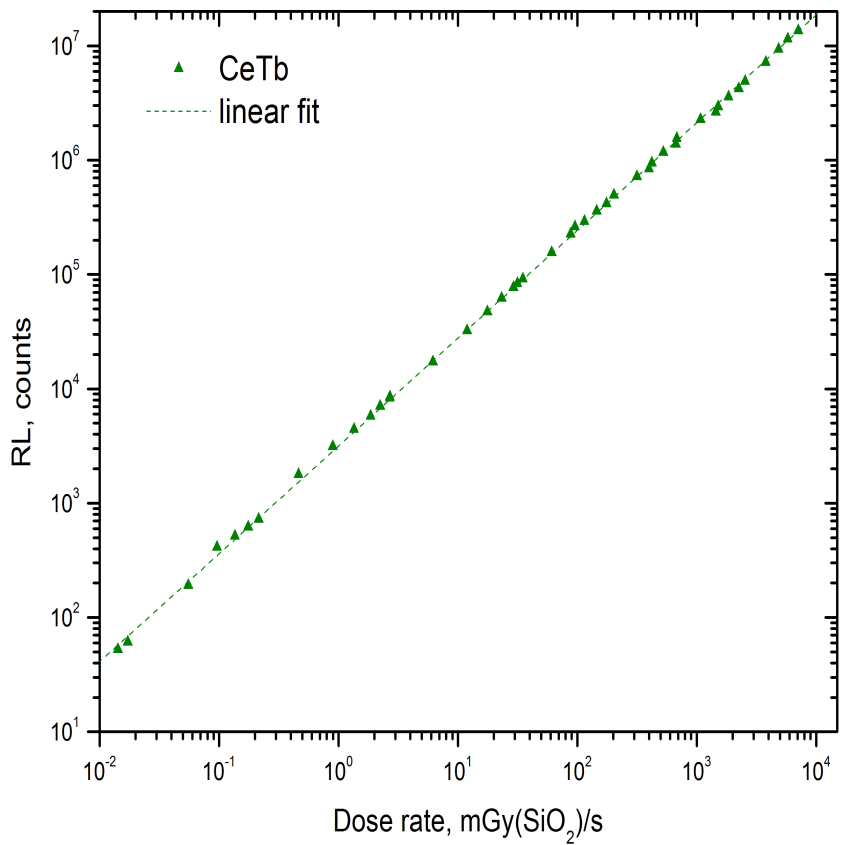

**Figure 7b**. Dose-rate dependence for Ce/Tb co-doped rod